# The unusual minimum of sunspot cycle 23 a consequence of Sun's meridional plasma flow variations

Dibyendu Nandy[1], Andrés Muñoz-Jaramillo[2,3] & Petrus C. H. Martens[2,3]

[1]*Department of Physical Sciences, Indian Institute of Science Education and Research, Kolkata, Mohanpur 741252, West Bengal, India*

[2]*Department of Physics, Montana State University, Bozeman, MT 59717, USA*

[3]*Harvard-Smithsonian Center for Astrophysics, Cambridge, MA 02138, USA*

**Direct observations over the past four centuries[1] show that the number of sunspots observed on the Sun's surface vary periodically, going through successive maxima and minima. Following sunspot cycle 23, the Sun went into a prolonged minimum characterized by a very weak polar magnetic field[2,3] and an unusually large number of days without sunspots[4]. Sunspots are strongly magnetized regions[5] and are generated by a dynamo mechanism[6] which recreates the solar polar field mediated via plasma flows[7]. Here we report results from kinematic dynamo simulations which demonstrate that a fast meridional flow in the early half of a cycle, followed by a slower flow in the latter half, reproduces both the characteristics of the minimum of sunspot cycle 23 – a large number of spotless days and a relatively weak polar field. Our model predicts that, in general, very deep minima are associated with weak polar fields. Sunspots govern the solar radiative energy[8,9] and radio flux, and in conjunction with the polar field, modulates the solar wind, heliospheric open flux and consequently cosmic ray flux at Earth[3,10,11]; our results therefore provide a causal link between solar internal dynamics and the atypical values of these heliospheric forcing parameters during the recently concluded solar minimum.**

Our current state of understanding[6] is that solar magnetic fields are produced by a dynamo mechanism in the Sun's interior involving the recycling of the toroidal and poloidal magnetic field components. The poloidal component is stretched by differential rotation to generate strong toroidal flux tubes; the latter are unstable to magnetic buoyancy and rise through the solar convection zone to erupt as tilted bipolar sunspot pairs, which subsequently decay and disperse to regenerate the poloidal component of the magnetic field[12,13,14]. Dynamo models based on this idea have been successfully utilized in explaining various observed features of the solar cycle[6] including studying their predictability[15,16,17]. In a complementary approach, solar surface flux transport models are used to study in detail the contribution of surface flux transport processes to the solar polar field evolution (quantified as the radial-component of the poloidal field). Surface flux transport simulations indicate that the polar field strength at cycle minimum is determined by a combination of factors, including the flux and tilt angles of bipolar sunspot pairs and the amplitude and profile of meridional circulation and super-granular diffusion[2,3].

Analysis of the sunspot tilt-angle distribution of cycle 23 shows that the average tilt angle did not differ significantly from earlier cycles[2]. The amplitude of the super-granular diffusion coefficient is also not expected to change significantly from cycle to cycle. However, the axisymmetric meridional circulation of plasma[18] – which is observationally constrained only in the top 10% of the Sun and has an average poleward speed of 20 m/s there – is known to exhibit significant intra- and inter-cycle variation[19-22]. The equatorward counter-flow of the circulation near the base of the convection zone is coupled through mass-conservation to the poleward surface flow and therefore this return flow should also be variable. It is believed that this equatorward return flow of plasma plays a crucial role; it drives the equatorward migration of sunspots, determines the solar cycle period and the spatio-temporal distribution of sunspots[23,22,6]. We perform kinematic solar dynamo simulations to investigate whether internal meridional flow variations can produce deep minima between cycles in general, and in particular, explain the defining characteristics of the minimum of cycle 23 – comparatively weak dipolar field strength and an unusually long period without sunspots (Fig. 1).

We use a recently developed axisymmetric, kinematic solar dynamo model[24] to solve the evolution equations for the toroidal and poloidal components of the magnetic field (see online Supplementary Information). This model has been further refined using a buoyancy algorithm that incorporates a realistic representation of bipolar sunspot eruptions following the double-ring formalism[25,26] and qualitatively captures the surface-flux transport dynamics leading to solar polar field reversal (including the observed evolution of the radial component of the Sun's dipolar field)[7]. In order to explore the effect of changing meridional flows on the nature of solar minima, one needs to introduce fluctuations in the meridional flow. The large-scale meridional circulation in the solar interior is believed to be driven by Reynolds stresses and small temperature differences between the solar equator and poles; variations in the flows may be induced by changes in the driving forces, or through the feedback of magnetic fields[27]. The feedback is expected to be highest at solar maximum (polar field minimum), when the toroidal magnetic field in the solar interior is the strongest. We therefore perform dynamo simulations by randomly varying the meridional flow speed at solar cycle maximum between 15 m/s—30 m/s (with the same amplitude in both the hemispheres) and study its effect on the nature of solar cycle minima.

Our simulations extend over 210 sunspot cycles corresponding to 1860 solar years; for each of these simulated cycles we record the meridional circulation speed, the cycle-overlap (which includes the information on number of sunspot-less days) and the strength of the polar radial field at cycle minimum. Fig. 2 shows the sunspot butterfly diagram and surface radial field evolution over a selected 40 year slice of simulation. Here cycle to cycle variations (mediated by varying meridional flows) in the strength of the polar field at minimum and the structure of the sunspot butterfly diagram are clearly apparent, hinting that the number of spotless days during a minimum is governed by the overlap (or lack-thereof) of successive cycles.

We designate the minimum in activity following a given sunspot cycle, say n, as the minimum of n (because the sunspot eruptions from cycle n contribute to the nature of this minimum). We denote the amplitude of the meridional flow speed after the random change at the maximum of cycle n as $v_n$, which remains constant through the minimum of cycle n and changes again at the maximum of cycle n+1. According to this convention, the speed during the early (rising) half of cycle n would be $v_{n-1}$. To explore the relationship between the varying meridional flow, the polar field strength and cycle overlap, we generate statistical correlations between these quantities separately for the northern and southern solar hemispheres from our simulations over 210 sunspot cycles.

Surprisingly we find that there is no correlation between the flow speed at a given minimum (say, $v_n$), and cycle overlap (or the number of sunspot-less days) during that minimum, while the polar field strength at that minimum ($B_r$) is only moderately correlated with $v_n$ (Fig. 3-a,b). Since transport of magnetic flux by the meridional flow involves a finite time, it is likely that the characteristics of a given minimum could depend on the flow speed at an earlier time. We find that this is indeed the case (Fig. 3-c,d), with cycle overlap (or number of spotless-days) and the polar field strength at a given minimum n, being strongly correlated with the flow speed $v_{n-1}$ (i.e., meridional flow during the early, rising part of cycle n). We also find that the cycle overlap is moderately correlated and the polar field strength ($B_r$) is strongly correlated with the change in flow speed from the earlier to the latter half of the cycle (Fig. 3-e,f). Taken together these results show that a fast flow during the early rising part of the cycle, followed by a relatively slower flow during the latter declining part of the cycle, results in a deep solar minimum.

The main characteristics of the minimum of solar cycle 23 are a large number of spotless days and relatively weak polar field strength. In Fig. 4 we plot the polar field versus cycle overlap and find that very deep minima are in fact associated with relatively weak polar field strengths. Thus, the qualitative characteristics of the unusual minimum of sunspot cycle 23 are self-consistently explained in our simulations driven by changes in the Sun's meridional plasma flow. Our model predicts that in general, extremely deep solar minima – with a large number of spotless days – would also be characterized by relatively weak solar polar field strength.

We find that our model results are robust with respect to reasonable changes in the driving parameters. Simulations with continuous flow variations (as opposed to discrete changes), relatively higher magnetic diffusivity, and a different threshold for buoyant active region eruption – all yield qualitatively similar relationships between the nature of solar minima and flow speed variations (see online Supplementary Information).

Valuable insights to our simulation results may be gained by invoking the physics of meridional flow mediated magnetic flux transport. A faster flow ($v_{n-1}$), before and during the early half of a cycle n would sweep the poloidal field of the previous cycle quickly through the region of differential rotation responsible for toroidal field induction; this would allow less time for toroidal field amplification and hence result in a sunspot cycle (n) which is not too strong. The fast flow, followed by a slower flow during the latter half of cycle n – persisting to the early part of the next cycle (n+1) would also distance the two successive cycles (i.e., successive wings in the sunspot butterfly diagram), contributing to a higher number of sunspot-less days during the intervening minimum. Moreover, a strong flow during the early half of cycle n would sweep both the positive and negative polarity sunspots of cycle n (erupting at mid-high latitudes) to the polar regions; therefore lower net flux would be available for cancelling the old cycle polar field and building the field of the new cycle – resulting in a relatively weak polar field strength at the minimum of cycle n. We believe that a combination of these effects contribute to the occurrence of deep minima such as that of solar cycle 23.

Independent efforts utilizing surface flux transport simulations show that surface meridional flow variations alone (observed during solar cycle 23, see also online Supplementary Information) is inadequate in reproducing the weak polar field of cycle 23[28]. Dynamo simulations – which encompass the entire solar convection zone – are therefore invaluable for probing the internal processes that govern the dynamics of the solar magnetic cycle, including the origin of deep minima such as that of cycle 23. We anticipate that the recently launched Solar Dynamics Observatory will provide more precise constraints on the structure of the plasma flows deep down in the solar interior, which could be useful for complementing these simulations.

**Supplementary Information** accompanies the paper on **www.nature.com/nature**.

**Acknowledgements** This work was supported through the Ramanujan Fellowship of the Government of India at the Indian Institute of Science Education and Research, Kolkata and by a NASA Living With a Star grant NNX08AW53G to the Smithsonian Astrophysical Observatory and Montana State University. We are grateful to Dr. David Hathaway for providing the observational data on sunspot-less days, whose analysis is reported in Fig. 1.

**Author Contributions** D.N. conceived the principal idea and in conjunction with P.C.H.M. and A.M.J. planned the simulations experiments – which were performed by A.M.J. under their guidance. D.N. provided the lead in interpreting the results and all three contributed to writing the paper.

**Author Information** The authors declare no competing financial interests. Correspondence and requests for materials should be addressed to D.N. (dnandi@iiserkol.ac.in).

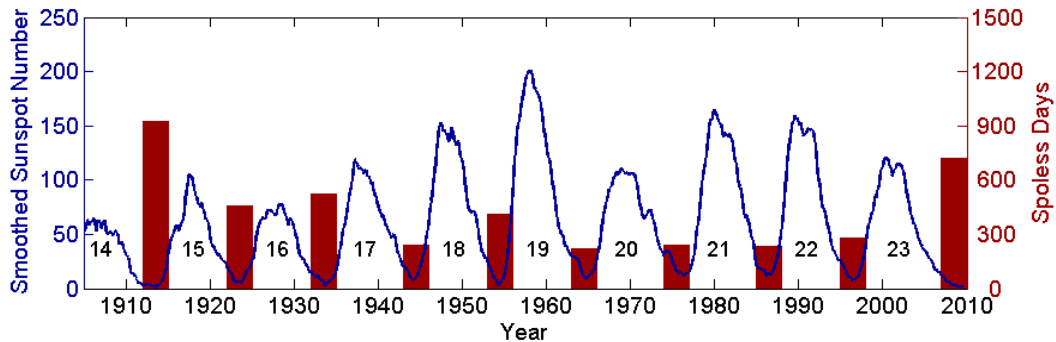

Fig.1: Sunspot observations over the past century: Starting with the pioneering telescopic observations of Galileo Galilei and Christopher Scheiner in the early 17$^{th}$ Century, sunspots have been observed more or less continuously up to the present. Except for a period during1645-1715 AD known as the Maunder minimum[29] when hardly any sunspots were observed, the sunspot time series shows a cyclic variation with an average period of 11 years going through successive epochs of maximum and minimum activity. The blue curve in the above figure shows this cyclic activity (the left-hand y-axis depicts sunspot numbers) over the past century. The period of minimum solar activity is often parameterized by the number of days without sunspots (i.e. spotless days). The red bars shows the cumulative number of sunspot-less days (right-hand y-axis) between successive maximum; the minimum of sunspot cycle 23 was the longest in the space age. Moreover, the solar polar magnetic field strength during this minimum was significantly weaker than the previous three minima for which direct polar magnetic field observations exists. Nonetheless, the recorded sunspot history shows solar cycles 13 and 14 had even larger number of spotless days; hence, although the recently concluded minimum was unusual, it is not unique. The explanation of such variations in the solar activity cycle should be sought in the physical mechanism that produces sunspots, i.e., solar dynamo theory.

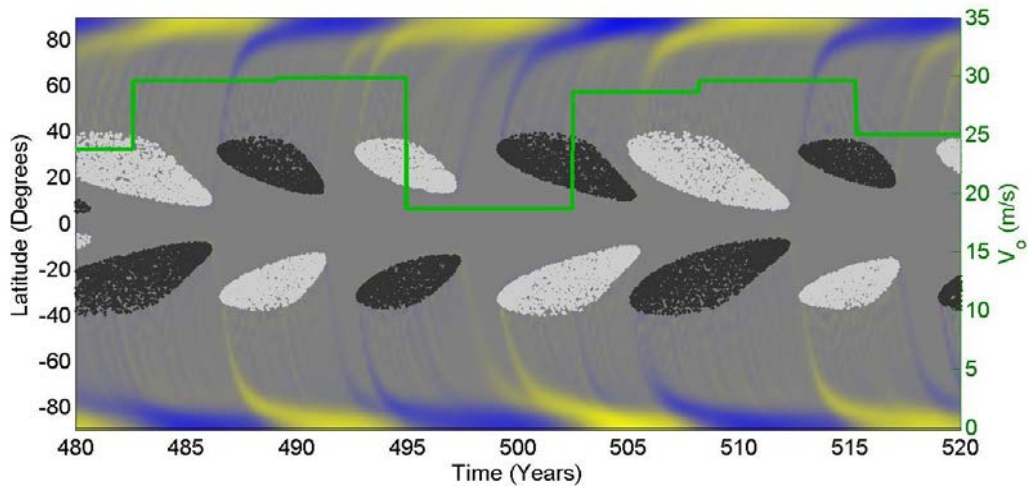

Fig.2: Simulated sunspot butterfly diagram from our solar dynamo simulations showing the time (x-axis)—latitude (left-hand y-axis) distribution of solar magnetic fields. The green line depicts the meridional flow speed which is made to vary randomly between 15—30 m/s (right-hand y-axis) at sunspot maximum, staying constant in between. The varying meridional flow induces cycle to cycle variations in both the amplitude as well as distribution of the toroidal field in the solar interior, from which bipolar sunspot pairs buoyantly erupt. This variation is reflected in the spatiotemporal distribution of sunspots shown here as shaded regions (lighter shade represents sunspots that have erupted from positive toroidal field and darker shade from negative toroidal field, respectively). The sunspot butterfly diagram shows varying degree of cycle overlap (of the "wings" of successive cycles) at cycle minimum. The polar radial field strength (depicted in colour, yellow-positive and blue-negative), is strongest at sunspot cycle minimum and varies significantly from one cycle minimum to another.

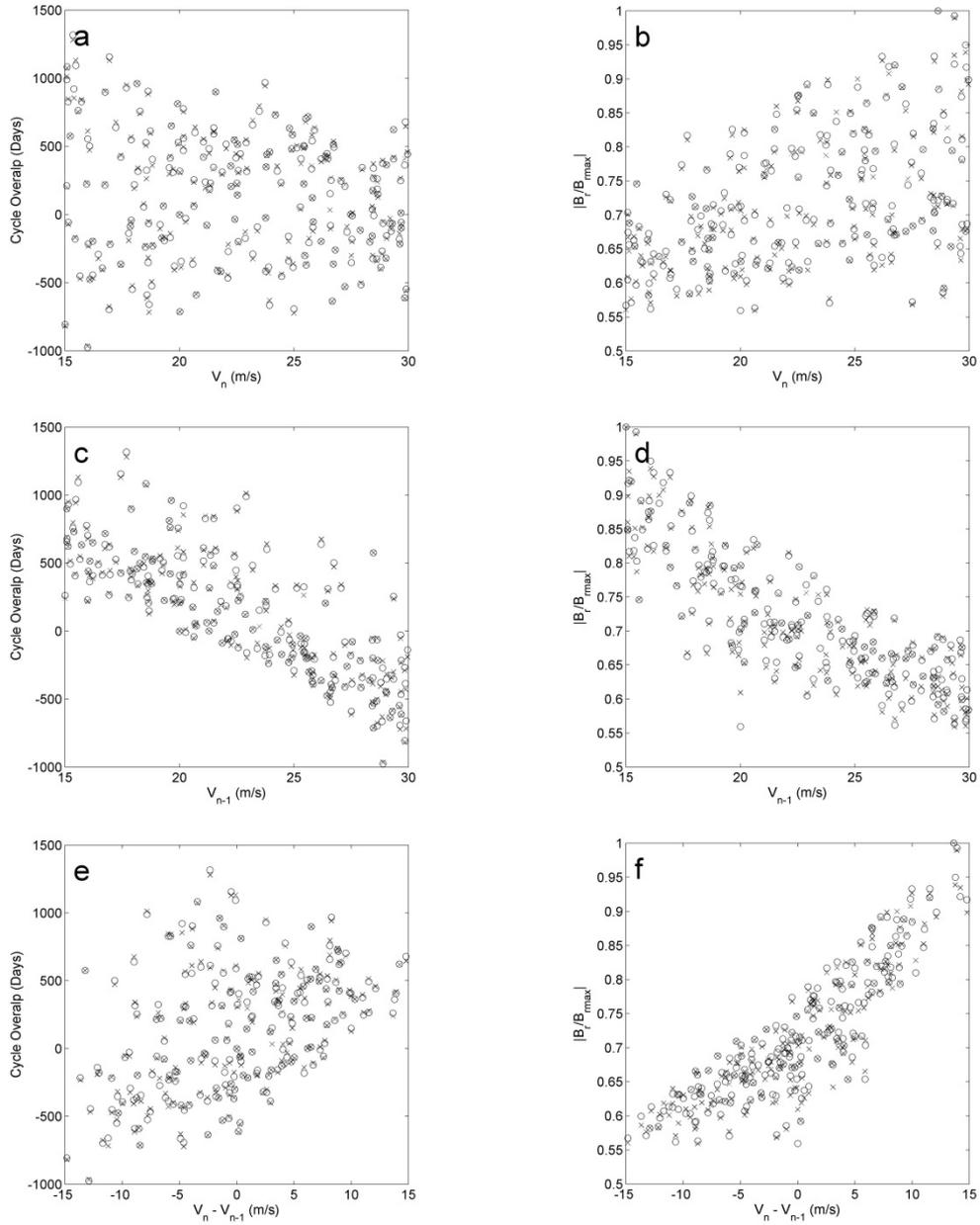

Fig. 3: Relationship between randomly varying meridional flow speed and simulated solar minimum characteristics quantified by cycle overlap and solar polar field strength. Here $v_n$ denotes flow speed during the minimum of sunspot cycle n, $v_{n-1}$ the speed during the early rising part of cycle n and $v_n - v_{n-1}$ the change in flow speed between the declining and the (early) rising part of the cycle. Cycle overlap is measured in days. Positive overlap denotes number of days where simulated sunspots from successive cycles erupted

together, while negative overlap denotes number of sunspot-less days during a solar minimum; large negative overlap implies a deep, (i.e., long) minimum. The polar field ($|B_r / B_{max}|$) is represented by the peak radial field attained during a solar minimum normalized with respect to the maximum radial field attained during the complete model run (here, $B_{max} = 16.66 \times 10^3$ Gauss; see online Supplementary Information for a discussion on polar field amplitudes). The relationship between the above parameters is determined by the Spearman's rank correlation coefficient (210 data points for each solar hemisphere, with northern and southern hemisphere data depicted as crosses and circles, respectively). A) Cycle overlap versus $v_n$; correlation coefficient for north (south) hemisphere: r = − 0.13 (− 0.13), confidence levels for north (south) hemisphere, p = 93.42% (94.53%). B) Polar field strength versus $v_n$: r = 0.45 (0.45), p = 99.99% (99.99%). C) Cycle overlap versus $v_{n-1}$: r = − 0.81 (− 0.80), p = 99.99% (99.99%). D) Polar field strength versus $v_{n-1}$: r = − 0.83 (− 0.83), p = 99.99% (99.99%). E) Cycle overlap versus $v_n - v_{n-1}$: r = 0.45 (0.45), p = 99.99% (99.99%). F) Polar field strength versus $v_n - v_{n-1}$: r = 0.87 (0.87), p = 99.99%, (99.99%). Evidently, a change from fast to slow internal meridional flow results in deep solar minima.

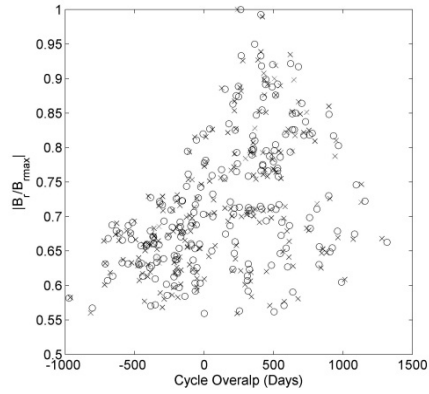

Fig. 4: Simulated normalized polar field (y-axis, similar convention as in Fig. 3) versus cycle overlap (x-axis) at sunspot cycle minimum in units of days. Spearman's rank correlation estimate: r = 0.46 (0.47), p = 99.99% (99.99%) for northern (southern) hemisphere data (depicted as crosses and circles, respectively). The results show that a deep solar minimum with a large number of spotless days is typically associated with a relatively weaker polar field – as observed during the minimum of sunspot cycle 23.